\documentclass[useAMS,usenatbib]{mn2e}

\usepackage{graphics}
\usepackage{epsfig}

\title[NGC 1023]{Extended Star Clusters in NGC 1023 from HST/ACS
Mosaic Imaging}
\author[D. A. Forbes et al.]{Duncan A. Forbes$^{1}$\thanks{E-mail:
dforbes@swin.edu.au}, Andres Almeida$^{2}$, Lee R. Spitler$^{3}$,
Vincenzo Pota$^{1}$ 
\\
$^{1}$Centre for Astrophysics \& Supercomputing, Swinburne
University, Hawthorn VIC 3122, Australia\\
$^{2}$Departamento de Ciencias Fisicas, Universidad Andres Bello,
Republica 220, Santiago, Chile\\
$^{3}$Department of Physics and Astronomy, Faculty of Sciences,
Macquarie University, Sydney, NSW 2109, Australia\\
$^{4}$Australian Astronomical Observatory, PO Box 296 Epping, NSW
1710, Australia\\
}
\begin{document}


\pagerange{\pageref{firstpage}--\pageref{lastpage}} \pubyear{2002}

\maketitle

\label{firstpage}

\begin{abstract}

Faint fuzzies are a relatively new class of star cluster,
first found in the nearby S0 galaxy NGC 1023 by Larsen \&
Brodie using WFPC2 images from the Hubble Space Telescope (HST). Here
we investigate the star cluster system of NGC 1023 using an eight
pointing mosaic of ACS images from HST. We identify blue and red
normal globular clusters (two of which are particularly luminous
and dense) and two ultra compact dwarf objects (with effective
radius $\sim$ 10
pc). With 
our more complete spatial coverage, we also find 81 red and 27 blue
faint fuzzies (FFs). We confirm the association of the red
FFs with the disk of NGC 1023, consistent with them being
long-lived open clusters. Half of the blue FFs
appear to be associated with the dwarf satellite galaxy NGC
1023A (which was largely absent from the original HST/WFPC2
coverage), while the remainder 
are spatially coincident with the densest HI 
gas that surrounds NGC 1023. 
The blue FFs have colours that are consistent with young (few 100
Myr old) star
clusters that formed during the most recent interaction between NGC 1023 and
its satellite NGC 1023A.

\end{abstract}

\begin{keywords}
galaxies: star clusters -- galaxies: evolution -- galaxies:
lenticular -- galaxies individual (NGC 1023)
\end{keywords}

\section{Introduction}

At the turn of century, Larsen \& Brodie (2000) reported the discovery of star
clusters in the nearby (11.1 Mpc) S0 galaxy NGC 1023 using
HST/WFPC2 imaging. As well as
normal globular clusters (see Kartha et al. 2014 for a recent
analysis), NGC 1023 hosts a population of faint objects with
large sizes (i.e. effective radii of 7-20 pc) and low surface
brightness. 
Most are red
in colour and subsequent follow-up spectroscopy with the Keck
telescope confirmed their high metallicity and old ages (Larsen
\& Brodie 2002). They dubbed these objects Faint Fuzzies
(FFs). They add to the ever increasingly family of old star
clusters which are filling the parameter space of size and
luminosity (e.g. Brodie et al. 2011; Bruens \& Kroupa 2012; 
Forbes et al. 2013).

Using the ACS VCS survey data, Peng et al. (2006) found a number
of similar objects in 12 early-type Virgo cluster galaxies (most
of which were morphologically classifed as S0). 
They suggested that the best Galactic
analog to these clusters, which they called 
Diffuse Star Clusters (DSCs), were old open clusters. 

Based on the kinematics of red FFs in NGC 1023, Burkert et al. (2005)
found them to lie in a fast rotating ring structure and suggested
that they had formed in a galaxy interaction. 
Chies-Santos et al. (2013) compared red FF kinematics with those of
Planetary Nebulae and the HI gas finding disk-like
kinematics and concluded that the red FFs are not associated with
an ongoing galaxy
merger but are simply long-lived open disk clusters. Further
support for the long-lived in-situ formation of the FFs comes
from the modelling of Br{\"u}ns et
al. (2009) who showed that merging star clusters could reproduce
well the size, mass and spatial distribution of FFs in NGC 1023. 


Thus FFs and DSCs appear to be old open clusters associated with disks in
lenticular galaxies. They may also be common in the disks of
late-type spirals but are difficult to detect (and therefore
study) within a complex,
dusty disk. 

Here we analyse a mosaic of eight HST/ACS images in two filters of NGC
1023. This is
one of only a few early-type galaxies with such extensive ACS 
spatial coverage.
It also includes the 
dwarf satellite galaxy NGC
1023A, which was largely missing from the HST/WFPC2 coverage. 
 With a wider and more continuous field-of-view  than the original study
of Larsen \& Brodie (2000) using HST/WFPC2 we search for additional red
FFs to confirm their disk-like 2D distribution. 
Larsen \& Brodie (2000) also found two {\it blue} FFs associated
with NGC 1023A.
Greater spatial
coverage will also aid in quantifying the blue FF population. 


\section{Data Reduction and Analysis} 

The eight HST/ACS F475W (g) and F850LP (z) band images used in this study 
were taken as part of proposal 12202 (PI: Sivakoff). Exposure
times were typically 768s in the g band and 1308s in the z band. 
The eight images create a rare ACS mosaic of a nearby early-type
galaxy, covering approximately 12 $\times$ 7 sq. arcmins. 

Objects were detected in the individual 
images using the SExtractor package (Bertins \& Arnouts 1996). Photometry
of these sources was carried out using DAOPHOT. After
the initial selection, 
magnitudes in 5 pixel (0.25 arcsec) radius aperture were measured and
an aperture correction applied based on isolated objects in each
image. Like the previous work of Larsen \& Brodie (2000), our
magnitudes may be systematically 
underestimated by a few tens of a magnitude for
the largest objects but the colours are largely independent of
any aperture correction. 
Total instrumental magnitudes were calibrated to the AB
photometric system using zero points of Sirianni et
al. (2005). Effective radii (which we refer to as 
R$_e$ or simply size) 
were measured using the ISHAPE code (Larsen
1999) with a King profile concentration parameter fixed to 30 (as is commonly
used, e.g. Larsen \& Brodie 2000; Usher et al. 2013). Thus R$_e$
is the half-light radius of the best fitting c = 30 King profile.
Common objects found in multiple pointings have their
magnitudes and errors averaged. 
Finally, we correct the magnitudes and colours for Galactic
extinction following Schlafly \& Finkbeiner (2011). Below we
quote extinction-corrected values. 

Star cluster candidates are
selected to have colours 0.6 $<$ g--z $<$ 1.8 (which corresponds
to --2.6 $<$ [Z/H] $<$ 1.1 according to the globular cluster
based transformation of Usher et al. 2012).
We also impose a faint magnitude cut of 
z $<$ 23.5, which is effectively a cut in photometric error of
less than $\pm$0.1. In addition, a minimum size of 0.3 pc was used to
effectively remove stars. We did not employ a maximum size limit
but a visual inspection was
carried out to conservatively remove background galaxies. Table 1 lists
the positions, g and z magnitudes, g--z colours and half-light effective
radii for the final 358 star cluster candidates. This includes a
dozen objects that may be associated with NGC 1023A.

\begin{table*}
\caption{Star clusters in NGC 1023. See online version for full table. 
}
\begin{tabular}{lccccccccccc}
\hline
ID & RA & Dec. & g & g$_{err}$ & z & z$_{err}$ & g--z
& g--z$_{err}$ & R$_e$ & R$_{err}$ & R$_{err}$\\ 
 & (J2000) & (J2000) & (mag) & (mag) & (mag) & (mag) & (mag) &
(mag) & (pc) & (pc) & (pc)\\
\hline
  137 &  40.0883 &  39.0554 &  24.199 &    0.090 &
22.952 &    0.109 &     1.246 &   0.141 &    0.37 &    1.01 &      -0.37\\   
  138 &  40.1050 &  39.0574 &  23.087 &    0.048 &
22.125 &   0.072 &     0.961 &   0.086 &    0.39 &   0.79 &     -0.32\\
  139  & 40.1041 &  39.0668 &  22.687 &   0.073 &    21.462
& 0.100  &   1.224 &   0.124 &    0.40 &    0.35 &    -0.40\\
 .. & .. & .. & .. & .. & .. & .. & .. & .. & .. & .. & ..\\
\hline
\end{tabular}
\\
Notes: Star cluster ID, Coordinates, g mag and error, z mag and
error, g--z colour and error, effective radius and errors.
\end{table*}

\section{Results and Discussion}

After applying the selection criteria described above, we show 
in Figure 1 the z
magnitude, g--z colour and effective radius in parsecs (assuming the objects
are associated with NGC 1023, i.e. at a distance of 11.1 Mpc). 

\begin{figure*}
\begin{center}
\includegraphics[scale=0.7,angle=-90]{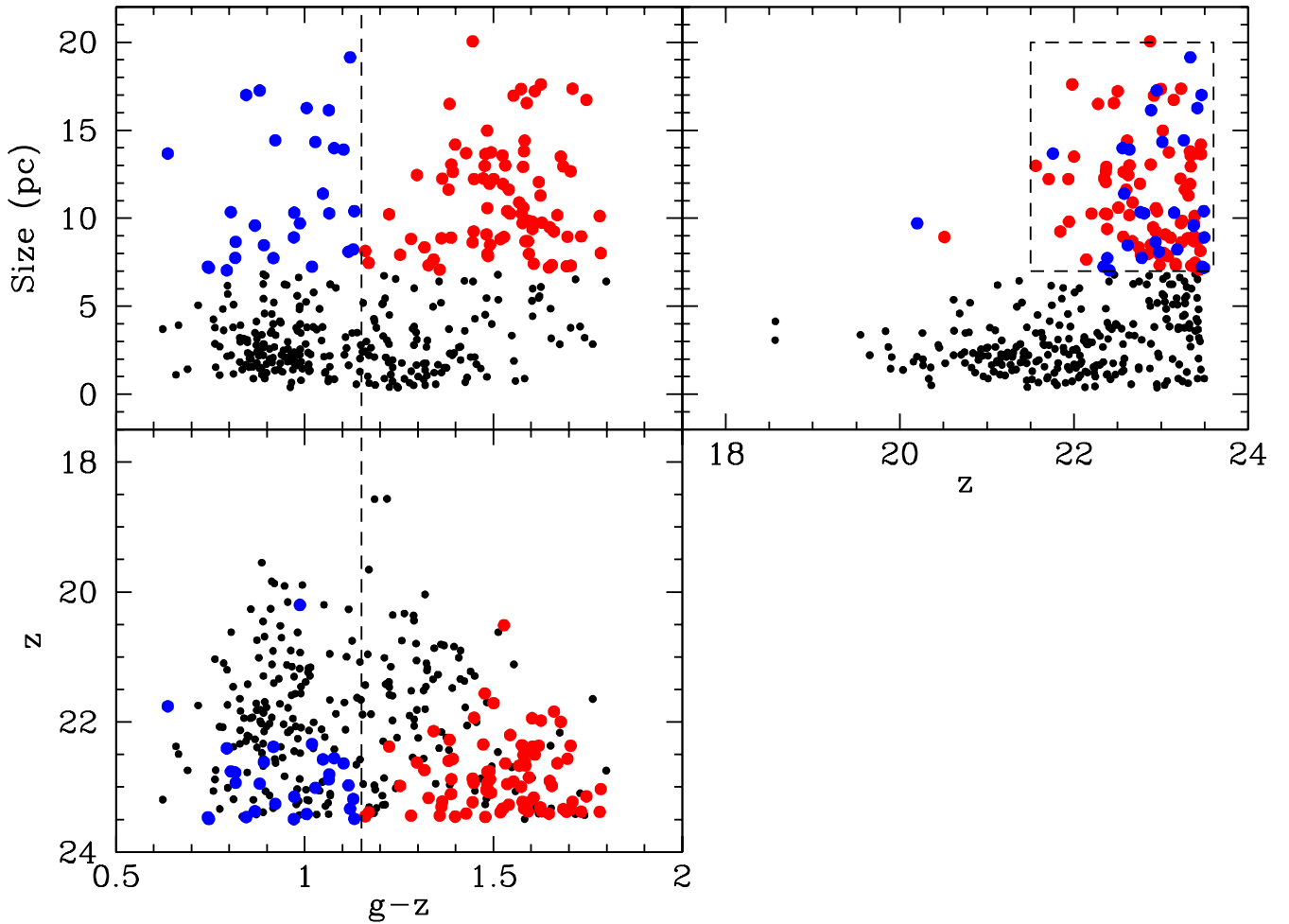}
\caption{Star cluster candidates in NGC 1023. In all panels,
  objects with sizes effective radii R$_e$
$>$7 pc are shown by larger symbols and colour-coded blue/red. {\it Top right} 
size in parsecs vs z magnitude. Note two
bright (z $\sim$ 18.5) compact globular clusters (GCs) and 2
ultra compact dwarfs (UCD) objects (size
$\sim$ 10 pc and z $\sim$ 20.5). A box shows the selection
criteria for Faint Fuzzies (FFs), 
i.e. R$_e$ $>$ 7 pc and 21.5 $<$ z $
<$ 23.5. Normal GCs have R$_e$ $\sim$ 3 pc. 
{\it Top left} size vs g--z colour. FFs are
divided into blue and red FFs at g--z = 1.15. {\it Lower left}
colour magnitude diagram. 
Two bright (z $\sim$ 18.5) GCs are clearly visible
with intermediate g--z colours.  
}
\end{center}
\end{figure*}

The size-magnitude diagram shows a locus of objects with 
R$_e$ $\sim$ 3 pc - these are globular cluster (GC) 
candidates. They include two bright
objects (IDs 33 and 269 with z $\sim$ 18.5, M$_z$ $\sim$ --11.73) 
that therefore have relatively high surface densities. In the
case of object 269 it has a size of R$_e$ = 4.1 pc and appears 
quite elongated. 
Both have spectra previously obtained from the Keck telescope
(Cortesi et al. 2014, in prep.)
confirming their association with NGC 1023, thus they appear to
be very luminous GCs. The object 269 
has been studied in detail by Larsen (2001) from HST/WFPC2
data. He measures the same size as we do and quotes an
ellipticity of 0.37. He notes that it is somewhat 
more compact than Omega Cen in the Milky
Way but similar in density to G1 in M31. The similarities with
Omega Cen and G1 suggest that 33 and 269 may both be the compact
nuclei of a stripped dwarf galaxy. 
We also note two relatively bright objects (IDs 102 and 315 with 
z $\sim$ 20.3, M$_z$ $\sim$ --9.7) with
large sizes (R$_e$ $\sim$ 10 pc) that resemble low luminosity
ultra compact dwarfs (UCDs; see Forbes et al. 2013). The blue UCD
(g--z = 0.99; ID: 102) has been confirmed with Keck spectroscopy (Cortesi et
al. 2014, in prep.) to be associated with NGC 1023. 
At faint magnitudes (z $>$ 21.5) a number of objects have sizes
$>$ 7 pc, i.e. FFs by the definition of Larsen \& Brodie (2000).
Using this selection 
we find 109 FF candidates, 16 of which are in
common with the original WFPC2 V and I band study of
Larsen \& Brodie (2000). Larsen \& Brodie found that the red FFs
do not have the standard Gaussian luminosity function of GCs, but
rather continued to grow in number down to their magnitude limit
of V $\sim$ 24. We find a similar behaviour for both the red and
blue FFs, From the study of a large sample of extended clusters
(R$_e$ $>$ 10 pc), Bruens \& Kroupa (2012) found that the peak or
turnover magnitude in early-type (i.e. both elliptical and
lenticular) galaxies occurred around M$_V$ = --6.4. If FFs have a
similar turnover magnitude (i.e. V $\sim$ 23.8 at the distance of
NGC 1023), then our data have reached the peak of the 
luminosity function of the FFs.

In the colour-magnitude diagram one can see the clear bimodality
of the two GC subpopulations, and at fainter magnitudes the FF
candidates. A simple colour cut between the blue and red GC
subpopulations at g--z = 1.15 is assigned. This corresponds to
[Z/H] $\sim$ --0.6 (Usher et al. (2012), and hence [Fe/H] $\sim$
--0.9. With this colour cut we find 81 red FFs and 27 blue FF
candidates (including 2 previously identified by Larsen \& Brodie
2000). 
This diagram also shows the location of the two UCDs and
the two high density GCs with intermediate colours.

The third panel shows the size-colour distribution. The FFs 
lie above the GCs in this panel with a similar distribution
in g--z colour (although the colour cut for the FFs could be redder by 0.05
mag. than for the GCs). 
Red FFs clearly outnumber the blue FFs (81 vs
27). Here we have followed the size cut of Larsen \& Brodie
(2000), i.e. at 7 pc. However, our selected star clusters hint at
three size regimes, i.e. normal GCs with R$_e$ $\sim$ 3 pc,
objects with sizes 3 $<$ R$_e$ $<$ 12 pc, and those with R$_e$
$>$ 12 pc (for the blue objects this may be R$_e$ $>$ 14 pc).
Larger numbers of objects are needed to verify if this
visual impression represents different types of star cluster. 

\begin{figure*}
\begin{center}
\includegraphics[scale=0.9,angle=0]{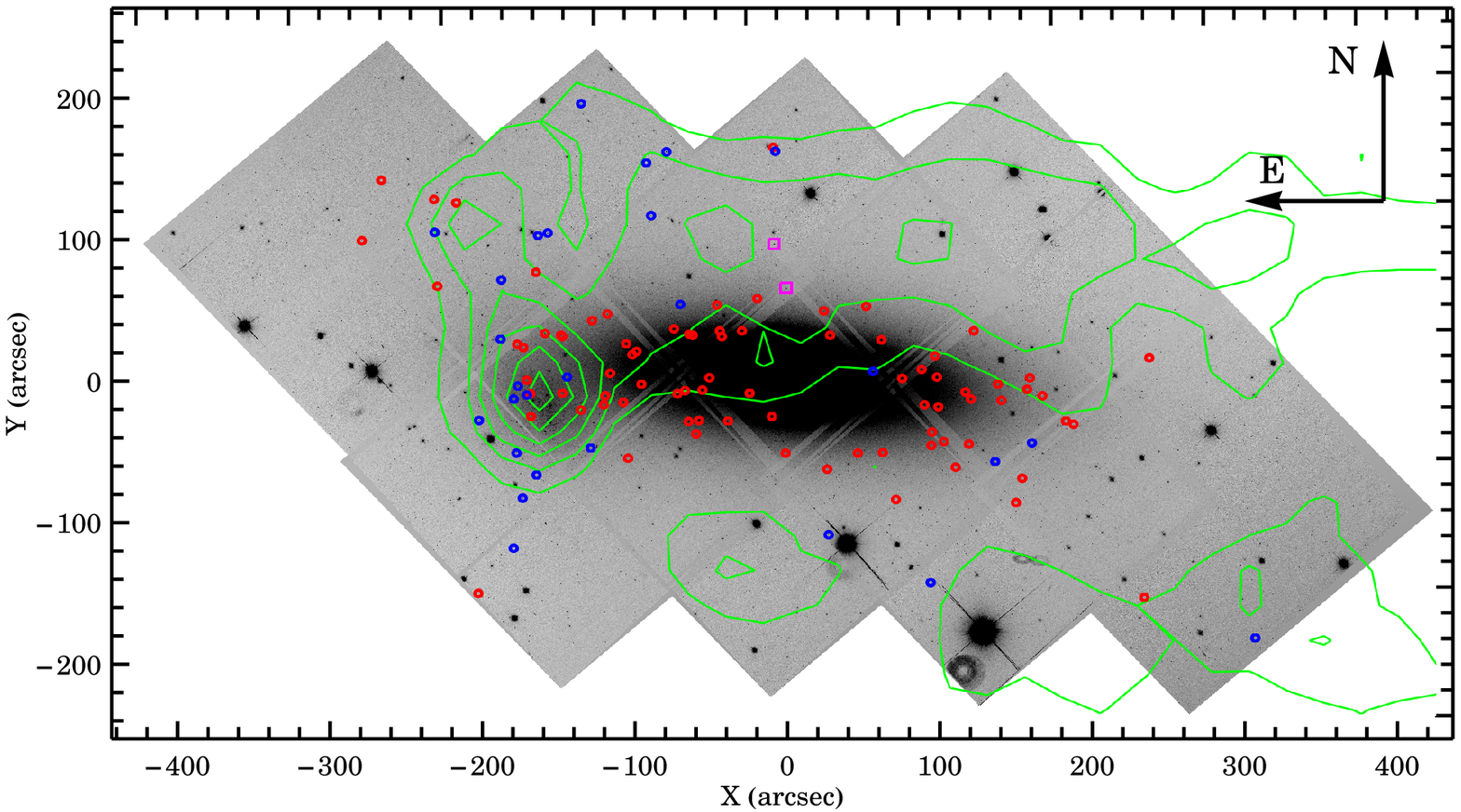}
\caption{HST/ACS mosaic of NGC 1023. 
The location of red FF (red circles) and blue FF (blue
circles) candidates are shown, along with two UCDs (magenta
squares).
The distribution of high density HI (from Morganti et al. 2006) 
is shown by the green contours.
The entire
field-of-view of the mosaic covers about 12 $\times$ 7 sq. arcmin.
The HI contours peak on the 
dwarf satellite, NGC 1023A, which can be seen at $\sim$ 140 arcsec East of NGC
1023 at PA $\sim$ 110$^o$. Photometry of star clusters was
carried out on individual pointings and not the mosaic shown
here. 
The red FFs follow the general distribution of the galaxy
disk, whereas roughly half of the blue FFs appear to be
associated with NGC 1023A and the others with the densest
HI gas. 
}
\end{center}
\end{figure*}

In Figure 2 we show the HST/ACS mosaic of NGC 1023, with the 
location of the blue and
red FFs, the two UCDs and the HI gas distribution (from Morganti et
al. 2006). 
Confirming previous works, the red FFs have a
distribution similar to that of the underlying stellar disk. 
The blue FFs do not follow the disk of NGC 1023. About half of them 
lie near NGC 1023A, while most of the others are coincident with
the densest HI gas. 

By fitting a simple model (i.e. equation 5 of Kartha et al. 2014)
for both the average ellipticity and position angle
simultaneously to the spatial distribution of red FFs, we find PA =
82 $\pm$ 3 degrees and b/a = 0.4 $\pm$ 0.2.
The stellar disk of NGC 1023 has an ellipticity of b/a = 0.26 and
PA of $\sim$85 degrees from a disk-bulge decomposition by Cortesi et
al. (2013). This supports the impression by eye in Figure 2, and
confirms the findings of Larsen \& Brodie (2000), that
red FFs are indeed associated with the disk of NGC 1023. In Figure 3 we show
the azimuthal distribution of the red and blue FFs, with the red
FFs peaking at the same PA as the galaxy major axis, while the
blue FFs peak at PA $\sim$ 110 degrees which is the direction of
NGC 1023A (while the distribution of other blue FFs shows no
clear peak in position angle). 


\begin{figure}
\begin{center}
\includegraphics[width=\columnwidth,scale=0.7,angle=0]{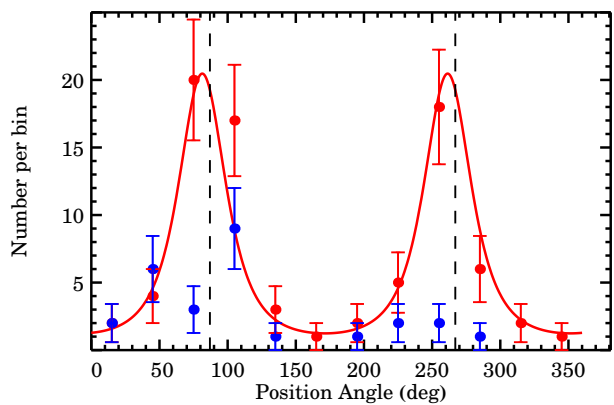}
\caption{Azimuthal distribution of faint fuzzy candidates in NGC
1023. The dashed lines represent the position angles (PAs) of the
major axis of NGC 1023, i.e. 85$^o$ and 265$^o$. Red FFs (red circles for
binned data and red line for a smooth model fit) 
show a peak at the same position angles as the
galaxy major axis, while the blue FFs (blue circles) 
peak PA $\sim$ 110$^o$ in the
direction of NGC 1023A. 
}
\end{center}
\end{figure}




The 27 blue and 81 red FFs have mean 
colours of g--z = 0.95 $\pm$ 0.03 and 
1.52 $\pm$ 0.01 respectively. 
In Figure 4 we show the mean colour of the blue and
red FFs compared to the single stellar population models of
Maraston (2005) with a Kroupa (2001) IMF.  
The red FFs have colours
associated with a metal-rich stellar population as expected if
they formed from disk material.
From Keck spectra the red FFs are known to be metal-rich and old
(Larsen \& Brodie 2002), which is consistent with Figure 4.

\begin{figure}
\begin{center}
\includegraphics[scale=0.5,angle=0]{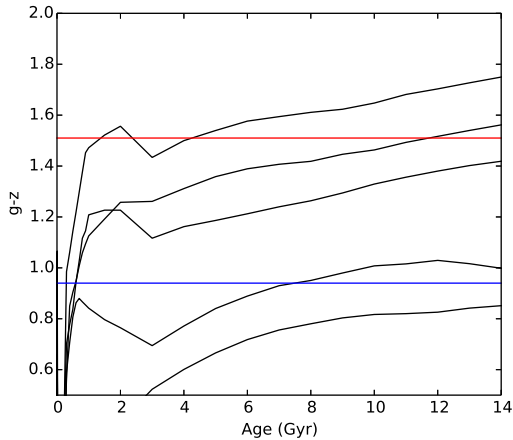}
\caption{Faint fuzzy mean colours compared to stellar population
models. The black solid lines show the evolution of g--z colour
with time for single stellar populations of different
metallicities (i.e. [Z/H] = --2.25, --1.35, --0.33, 0.00, +0.35
from bottom to top) from the models of Maraston (2005). 
The solid red and blue lines show the mean g--z colours of the
red and blue FFs respectively. The red FFs have colours
associated with a metal-rich stellar population as expected if
they formed from disk material; whereas the blue
FFs are consistent with being very young ($\le$ Gyr) as expected
if formed moderately enriched gas 
during the recent interaction with NGC 1023A. 
}
\end{center}
\end{figure}

The blue FFs are consistent with being very young ($\le$ Gyr). 
Half of the blue FFs appear to be
associated with NGC 1023A, which is thought to have interacted
with NGC 1023 a few 100 Myr ago and is responsible for the 
$\sim$10$^9$ M$_{\odot}$ of surrounding HI (Capaccioli et
al. 1986). Spectroscopy of two blue NGC 1023A 
star clusters have ages up to 500 Myrs (Larsen \& Brodie
2002), consistent with them forming during the 
interaction process. As the HI gas is widely
distributed around the NGC 1023 system the blue FFs not
coincident with NGC 1023A may also be a few 100 Myrs old
 -- followup spectroscopy is required. Thus
blue FFs may be the lower mass counterparts to tidal dwarf galaxies,
forming out of the tidal debris of a past galaxy interaction.

\section{Conclusions}

Using a recently-available mosaic of eight HST/ACS pointings in g
and z bands, we investigate the star cluster system
of NGC 1023 measuring magnitudes, colours and effective radii. 
We identify the normal globular cluster (GCs) system of
NGC 1023 with its red and blue subpopulations, and two
particularly luminous and therefore dense GCs. Two ultra compact
dwarf objects with effective radii $\sim$ 10 pc are also found.
In this Letter, we focus on the faint, extended 
(effective radii $>$ 7 pc) clusters called Faint Fuzzies
(FFs). Such objects were originally discovered in NGC 1023 by
Larsen \& Brodie (2000) using two HST/WFPC2 pointings. With our
more complete spatial coverage, we confirm the association of
some 81 red
FFs with the disk of NGC 1023 consistent with them being
long-lived open clusters (albeit somewhat larger and more
luminous than typical Milky Way open clusters). 
We also identify a population of 27 blue FFs, half of
which appear to be associated with the dwarf satellite galaxy NGC
1023A (which was largely absent from the original HST/WFPC2
coverage). The others appear to be spatially coincident with 
the HI gas that surrounds the NGC 1023 system.
The blue FFs are consistent with young (few 100 Myrs) star
clusters that formed during the interaction between the two
galaxies. 

\section{Acknowledgements}

We thank A. Cortesi, A. Chies-Santos and S. Larsen for useful
comments. 
We also the referee for some useful comments. 
Based on observations made with the NASA/ESA Hubble Space
Telescope, obtained from the data archive at the Space Telescope
Science Institute. STScI is operated by the Association of
Universities for Research in Astronomy, Inc. under NASA contract
NAS 5-26555.
DAF thanks the
ARC for financial support via DP130100388.

\section{References}

Bertin, E., Arnouts, S., 1996, A\&AS, 117, 393\\
Brodie J.~P., Romanowsky A.~J., Strader J., Forbes D.~A., 2011, AJ, 142, 
199 \\
Br{\"u}ns R.~C., Kroupa P., Fellhauer M., 2009, ApJ, 702, 1268 \\
Br{\"u}ns R.~C., Kroupa P., 2012, A\&A, 547, A65 \\
Burkert A., Brodie J., Larsen S., 2005, ApJ, 628, 231\\
Capaccioli M., Lorenz H., Afanasjev V.~L., 1986, A\&A, 169, 54\\
Chies-Santos A.~L., Santiago B.~X., Pastoriza M.~G., 2007, A\&A, 467, 1003 \\
Chies-Santos A.~L., Cortesi A., Fantin D.~S.~M., Merrifield
M.~R., Bamford S., Serra P., 2013, A\&A, 559, A67 \\
Cortesi A., et al., 2013, MNRAS, 432, 1010 \\
Forbes D.~A., Pota V., Usher C., Strader J., Romanowsky A.~J., Brodie 
J.~P., Arnold J.~A., Spitler L.~R., 2013, MNRAS, 435, L6 \\
Kartha S.~S., Forbes D.~A., Spitler L.~R., Romanowsky A.~J., Arnold J.~A., 
Brodie J.~P., 2014, MNRAS, 437, 273 \\
Kroupa P., 2001, MNRAS, 322, 231\\
Larsen S.~S., 1999, A\&AS, 139, 393\\
Larsen S.~S., 2001, AJ, 122, 1782\\
Larsen S.~S., Brodie J.~P., 2000, AJ, 120, 2938 \\
Larsen S.~S., Brodie J.~P., 2002, AJ, 123, 1488 \\
Maraston C., 2005, MNRAS, 362, 799 \\
Morganti R., et al., 2006, MNRAS, 371, 157 \\
Peng E.~W., et al., 2006, ApJ, 639, 838 \\
Schlafly E.~F., Finkbeiner D.~P., 2011, ApJ, 737, 103\\
Sirianni M., et al., 2005, PASP, 117, 1049 \\
Usher C., et al., 2012, MNRAS, 426, 1475\\
Usher C., Forbes D.~A., Spitler L.~R., Brodie J.~P., Romanowsky A.~J., 
Strader J., Woodley K.~A., 2013, MNRAS, 436, 1172 \\

\end{document}